\documentclass[10pt]{elsart}

\usepackage{graphicx}
\usepackage{epsf}
\usepackage{wrapfig}
\usepackage{here}

\begin{document}
\begin{frontmatter}

\title
{On design studies for the future 50 GeV arrays of imaging air 
\v{C}erenkov telescopes}

\author{Alexander K. Konopelko}

\address
{Max-Planck-Institut f\"ur Kernphysik, D-69029 
Heidelberg, Germany}

\begin{abstract}
{Arrays of imaging air \v{C}erenkov telescopes (IACTs) like VERITAS, HESS have 
been recently proposed as the instruments of the next generation for ground 
based very high energy $\gamma$-ray astronomy invading into 50-100 GeV energy 
range. Here we present results of design studies for the future IACT arrays  
which have been performed by means of Monte Carlo simulations. \\
We studied different trigger strategies, abilities of cosmic ray rejection for  
arrays of 4 and 16 telescopes with 10 m reflectors, equipped with cameras 
comprising 271 and 721 pixels of $0.25^\circ$ and $0.15^\circ$, respectively. 
The comparative analysis of the performance of such telescope arrays has been 
done for both camera options, providing almost the same field of view of 
$\sim 4.3^\circ$. \\
An important issue is the choice of the optimum {\it spacing} between the 
telescopes in such an array. In order to maximize the signal-to-noise ratio in 
observations at the small zenith angles of $\sim20^\circ$ as well as at large 
zenith angles of $\sim 60^\circ$, different arrangements of IACT array have been 
examined. Finally, we present a major recommendations regarding the optimum 
configuration.

\vspace{2 mm}


}
\end{abstract}

\end{frontmatter}

\noindent
{\bf 1. Introduction}

Presently the Very High Energy (VHE) domain, from 300 GeV to 20 TeV, of the cosmic 
$\gamma$-rays is covered by ground based instruments. Imaging air \v{C}erenkov 
telescopes (IACTs) are able to detect signals from galactic and extragalactic 
$\gamma$-ray emitters within one hour of observations and to measure their energy 
spectra with a few hours of good data. The variety of the physics results obtained 
with the currently operating IACTs as well as physics considerations for the 
forthcoming instruments (Weekes et al., 1997) suggest a prosperous future of VHE 
$\gamma$-ray astronomy indeed. \\
A major trend in the development of this technique is towards {\it stereoscopic 
arrays} of 10 m IACTs, such as the VERITAS (e.g., Weekes, 1997) and HESS (e.g., 
Hofmann, 1997) projects, approaching an energy threshold of $\sim$50-100 GeV. An 
alternative approach might be the construction of a single large (17 m) imaging 
air \v{C}erenkov telescope (MAGIC) in order to achieve the energy threshold as low 
as 20~GeV (Lorenz, 1997). The stereoscopic imaging of $\gamma$-ray-induced air 
showers has several advantages compared with a stand alone telescope (see Aharonian 
\& Konopelko, 1997) which provide high quality $\gamma$-ray observations. \\
The design of the future IACTs arrays is constrained, first of all, by the {\it 
physics goals}: we are interested in detection of the $\gamma$-ray sources at large 
distances (redshifts: $\rm z \geq 0.1$) with presumably low fluxes. To enlarge the 
number of targets it is also worth to perform the surveys. Future observations of 
AGNs would in addition need a good energy resolution over a broad energy range as 
well as an ability of long time monitoring of a few sources simultaneously. 
Observations of the extended, diffuse $\gamma$-ray sources (e.g., SNRs) would finally 
need a large field of view in order to accommodate objects of $\sim 0.5-1^\circ$ 
angular size (V\"{o}lk, 1993). For all that one can define desirable {\it physics 
parameters} of an IACT array: an effective energy threshold of $\sim 50-100$~GeV, a 
sensitivity to $\gamma$-ray fluxes as low as $\rm J_\gamma (> \, 100\, GeV) \simeq 
10^{-11} \, cm^{-2} s^{-1}$, an energy dynamic range up to 50 TeV, an angular 
resolution of $0.1^\circ$, an energy resolution of $\leq$20~\%, and a relatively 
large field of view ($\geq 4^\circ$). \\
Our previous studies (Aharonian et al., 1997) have shown that these physics parameters 
could be provided with an array of $\sim 10$ m telescopes, placed at $2.2$ km height 
above sea level. However, the design of the camera (e.g., the angular size of the 
pixels and the camera field of view) as well as the layout of the telescopes have 
still to be optimized. For that Monte Carlo simulations have been done. The major 
results of the simulations are summarized here.
 
\noindent
{\bf 2. Simulations}

The ALTAI computational code was used for generating $\gamma$-ray- and proton-induced 
air showers in the energy range 10~GeV - 50~TeV for two zenith angles of $20^\circ$ 
and $60^\circ$. For each shower the response of an extended array of IACTs, arranged 
in a rectangular lattice of 1000 x 600 $\rm m^2$ with  33 m step (589 nodes in total) 
was saved. The calculations have been done for 10 m ($\rm S \simeq$ 82$\rm \, m^2$) 
telescopes equipped by a camera in two alternative designs: (1) 271 pixels of $0.25^\circ$; 
(2) 721 pixels of $0.15^\circ$. Both designs give almost the same field of view of 
$4.3^\circ$. The mean night sky content per $0.15^\circ$ pixel per trigger time gate 
of 10~ns and photon-to-photoelectron conversion efficiency of 0.1 was taken as 1 ph.e.    
  
\noindent
{\bf 3. Results and discussion}

The sensitivity of IACT technique is determined by the effective detection area of the 
$\gamma$-rays, and the ability of cosmic ray rejection using the orientation and shape 
parameters of the \v{C}erenkov light images. The resulting signal-to-noise ratio for 1~hr 
observations is given by 
$\rm S/N = \eta^{(o)} \eta^{(s)}A_\gamma J_\gamma{(A_{cr} J_{cr})^{-1/2}} {t}^{1/2}$
where $\rm A_\gamma, A_{cr}$ are the detection areas for the $\gamma$-rays and cosmic rays, 
respectively, with the corresponding fluxes 
$\rm J_\gamma, J_{cr}$, and $\eta^{(o,s)} = \kappa^{(o,s)}_\gamma / (\kappa^{(o,s)}_{cr})^{1/2}$ 
are the enhancement factors after application of the orientation and shape analysis cuts 
($\gamma$-ray selection criteria). In general, the optimum design of the IACT array should 
give the maximum S/N ratio. The maximum $\gamma$-ray detection area mainly depends on the 
system trigger scheme and threshold as well as on the telescope arrangement (basically on 
the distance, $l$, between neighbour telescopes) whereas $\eta^{(o,s)}$ depend on the 
camera pixellation and field of view, and on the telescope spacing in the array. Note that 
all these relations are strong functions of the primary $\gamma$-ray energy. Using the 
Monte Carlo simulations for the dense grid of the telescopes the optimization of S/N ratio 
is straightforward. In the following discussion we try to disentangle most important 
relations in order to make clear the results of a complete array optimization. 
 
{\bf Gamma-ray detection rate.} The background light per pixel per trigger time gate ($\sim 10$ ns) 
sets the minimum trigger threshold at the level of 8 and 13 ph.e. for the local trigger schemes: 
3/721 (signal in each of any three neighbour pixels from 721 exceed the trigger threshold) and 
2/271 for two camera designs, respectively. The global system trigger demands at least two telescopes 
to be triggered locally within the time gate of 50~ns. It limits the random noise trigger rate at 
$0.1 \, \rm Hz$ with a corresponding single-telescope trigger rate of $\sim 400$ Hz (W. Hofmann, 
private communications). For these conditions both camera designs give $\sim$50~GeV energy threshold, 
determined as the energy corresponding to the maximum of the differential $\gamma$-ray detection rate, 
assuming the $\gamma$-ray energy spectrum $\rm dN_\gamma/dE \propto E^{-2.5}$. The integral $\gamma$-ray  
rate, $\rm R_\gamma$, at zenith, is expected to be about 1 Hz from the Crab Nebula for a 4 IACTs 
array and remains almost constant for a telescope spacing of less that $\sim$~120~m. Note that after 
the conventional software analysis cuts the number of survived low energy $\gamma$-rays of $\sim$50~GeV 
is noticeably reduced ($\geq 30$\%) and the peak in the differential detection rate may shift to the 
higher energies ($\sim 100$~GeV) depending on the assumed $\gamma$-ray spectrum.  

{\bf Angular resolution.} The stereoscopic observations allow to measure the shower direction by 
superposition of a several images in one common focal plane. The accuracy of this reconstruction -- 
the angular resolution -- can be defined as one standard deviation of the difference between the 
true and reconstructed direction of the $\gamma$-ray showers. Our Monte Carlo simulations show that 
the angular resolution for the array of telescopes in two camera designs, noted above, is roughly 
the same for small zenith angles ($\sim 20^\circ$). The angular resolution strongly depend on the 
average content of the background light in the camera pixels. A fine pixellation ($0.15^\circ$) helps 
to improve drastically the angular resolution at large zenith angles ($60^\circ$). The angular resolution 
of the telescope array substantially improves by increasing the baseline distance, l, up to $\sim$~120~m 
for $20^\circ$ zenith angle and up to $\sim 360$ m at $60^\circ$ zenith angle. For the energy threshold 
of 50~GeV, the optimum layout gives $\eta^{(o)} \simeq 3$ with 70 \% of the $\gamma$-rays within 
$0.3^\circ$ for the small zenith angles.  

{\bf Cosmic ray rejection.} For the second moment analysis the Monte Carlo simulations do not show 
noticeable difference in the ability of cosmic ray rejection for the two camera options. Using the 
{\it mean scaled Width} parameter one can get $\eta^{(s)} \sim 3$ for an energy threshold of 50~GeV. Note 
that $\eta^{(s)}$ does not depend on the telescope spacing whereas it substantially increases at 
energies far above the energy threshold ($\geq 100$~GeV). Multi-telescope coincidences (3,4) give 
better cosmic ray rejection. The low energy $\gamma$-ray triggers $\sim 50$~GeV provide a dominant 
rate of 2 and 3 pixel events for the camera designs of a coarse pixellation ($0.^\circ 25$). In this 
case the ratio of a minimum pixel signal to the trigger threshold is close to one and these events are 
unlikely to be extracted from the cosmic rays. Note that for good imaging the minimum pixel signal 
(slightly above the image tail cut) has to be substantially lower than the telescope trigger threshold 
providing the accurate measure of the image shape by sufficient number of pixels. 
 
{\bf Field of view}. The images of low energy $\gamma$-rays ($\rm E_\gamma \leq 100$~GeV) are 
concentrated very close to the camera center ($\sim 0.5^\circ$). The \v{C}erenkov light images from air 
showers observed at large zenith angles also shrink to the camera center because of a large distance from 
the shower maximum to the telescope. In observations at small zenith angles ($\sim 20^\circ$) the high 
energy showers ($\rm E \geq 10$ TeV) are partially truncated by the camera edge. The effective detection 
of such events needs an extended camera field of view up to at least $\sim4.5^\circ$. A large field of view 
is important for the observations of diffuse sources as well as for performing large area surveys. 

{\bf Optimum baseline distance.} We have tested different possible arrangements of 16 IACTs. 
For instance one can set the array layout as a sparse grid, or, as a 4 independent cells (which are scattered 
on the observation plane at the distances $\geq 250$~m) with 4 telescopes within each cell. Our simulations 
show that both layouts give almost the same integral $\gamma$-ray detection rates and final signal-to-noise 
ratio, whereas the grid structure is preferable for the registration of the low energy $\gamma$-rays 
($\rm E\sim 50-100\, GeV$). The resulting S/N ratio strongly depends on the {\it spacing} (grid baseline 
distance) between the telescopes. The optimum baseline distance for observations at small zenith angles 
($20^\circ$) is about 120 m. A further increase of the baseline distance leads to the corresponding increase 
in the energy threshold of the system. Large zenith angle observations need a large distance between the 
telescopes (up to 360 m). For that the peripheral telescopes of a grid can be effectively used.    

\noindent
{\bf Summary}

The physics motivations and current design studies allow us to set the possible arrangement of a low energy 
($\geq 50$~GeV) IACT array as a quadrangular grid of 16 telescopes with $\sim$120~m baseline distance. 
Each of telescopes has a 10 m reflector and is equipped with a fine resolution camera: 
817 pixels of $0.16^\circ$, which covers $\sim 4.5^\circ$ field of view (optionally a number of 
telescopes could have a camera of a large field of view $\geq5^\circ$ with the pixels of $0.25^\circ$). 
The array permits different operational modes, which include the observations with a complete array 
(search at the level of a maximum sensitivity), the large zenith angle observations, search for diffuse, 
extended $\gamma$-ray sources as well as performing large area surveys. Such an array allows monitoring 
of several $\gamma$-ray sources with the subgroups of 4 telescopes in order to maintain simultaneously a 
number of research programs.    
     
\noindent
{\bf References}

{\small

Aharonian F., Hofmann W., Konopelko A., V\"{o}lk H., 1997, {\it Astroparticle Physics}, 6, 343; 369 \\
Aharonian F., Konopelko A., 1997, {\it Proc. Workshop on TeV $\gamma$-ray Astrophys.}, South Africa, 
Ed. O.C. de Jager. 263\\ 
Hofmann W.,  {\it Proc. Workshop on TeV $\gamma$-ray Astrophys.}, South Africa, Ed. O.C. de Jager. 405\\
Lorenz E., 1997, {\it Proc. Workshop on TeV $\gamma$-ray Astrophys.}, South Africa, Ed. O.C. de Jager. 415 \\
V\"{o}lk H.J., 1993, {\it Proc. Workshop on TeV $\gamma$-ray Astropart.}, Calgary, Edt. R.C. Lamb, 32 \\      
Weekes T., et al., 1997, {\it Proc. 4th Compton Sym.}, Part 1, Williamsburg, 361 \\
Weekes T., 1997, {\it Proc. Workshop on TeV $\gamma$-ray Astrophys.}, South Africa, Ed. O.C. de Jager. 433
}
\end{document}